\documentclass{article}
\usepackage{amsmath}
\usepackage{algorithm2e}
\usepackage{multicol}
\usepackage{enumerate}
\usepackage{graphicx}
\usepackage{caption}
\usepackage{subcaption}
\newcommand*{\rom}[1]{\expandafter{\romannumeral #1\relax}}
\usepackage[rgb,dvipsnames]{xcolor}
\usepackage{lipsum}
 \usepackage[normalem]{ulem} 

\DeclareCaptionLabelSeparator{periodspace}{.\quad}
\begin{document}
\title{ \Large{ \textbf{Exploring the Relation Between Two Levels of Scheduling Using a Novel Simulation Approach}}}
\date{}




\author{Ahmed Eleliemy, Ali Mohammed, and Florina M. Ciorba\\
	Department of Mathematics and Computer Science\\
		University of Basel, Switzerland}

\date{}
\maketitle
\clearpage
\tableofcontents
\clearpage


%


\maketitle
\sloppy

\begin{abstract}
	
Modern high performance computing (HPC) systems exhibit a rapid growth in size, both ``horizontally" in the number of nodes, as well as ``vertically" in the number of cores per node. 
As such, they offer additional levels of hardware parallelism.
Each such level requires and employs algorithms for appropriately scheduling the computational work at the respective level. 
The present work explores the relation between two scheduling levels: batch and application. 
Understanding this relation is important for improving the performance of scientific applications, that are scheduled and executed in batches on HPC systems.	
The relation between batch and application level scheduling is understudied in the literature.
Understanding the relation and interaction between these two scheduling levels requires their simultaneous analysis during operation. 
In this work, such an analysis is performed via simultaneous simulation of batch and application level scheduling for a number of scenarios.
A generic simulation approach is presented that bridges two existing simulators from the two scheduling levels.
A novel two-level simulator that implements the proposed approach is introduced. 
The two-level simulator is used to simulate all combinations of three batch scheduling and four application scheduling algorithms from the literature. 
These combinations are considered for allocating resources and executing the parallel jobs of two batches from two production HPC systems.
The results of the scheduling experiments reveal the strong relation between the two scheduling levels and their mutual influence.
Complementing the simulations,
the two-level simulator produces standard parallel execution traces, which can visually be examined and which illustrate the execution of different jobs and, for each job, the execution of its tasks at node and core levels, respectively.
\end{abstract}

\paragraph*{keywords}
	
High performance computing; 
Batch level scheduling; 
Application level scheduling; 
Two-level scheduling; 
GridSim; 
Alea; 
SimGrid; 
SimDag; 
OTF2; 
Vampir.

\newpage
\section{Introduction}
\label{sec:intro}

In \mbox{cluster-based} HPC systems, the job scheduling and the resource management are not only key factors to achieving maximum cluster utilization.
They are also key factors to achieving user satisfaction in terms of job execution time and job waiting time~\cite{ahn2014flux}.
The job scheduling problem is well-known to be \mbox{NP-Complete~\cite{NpCompleteScheduling}}. 
Therefore, in the scheduling literature, a large number of job scheduling heuristics have been proposed.
Each of these heuristics has its own strengths in terms of achieved performance. 
The job scheduling policies include traditional algorithms, such as First~Come~First~Serve~(FCFS), Earliest~Deadline~First~(EDF), and Shortest~Job~First~(SJF)~\cite{azmi2011scheduling}, as well as \mbox{non-traditional} scheduling algorithms, implemented in commercial software, such as PBS~\cite{PBS} and LSF~\cite{LSF}.

The batch scheduling algorithms are no longer sufficient to exploit \emph{alone} all available parallelism of modern and future HPC systems. 
Due to the rapid developments in HPC technology and system design, HPC systems are growing along two orthogonal dimensions: sheer number of processing elements and complexity of the processing elements. 
Modern HPC systems combine hundreds of thousands and even millions of processing elements. 
The processing elements themselves are also no longer of simple architectures, as each can combine multiple CPUs, each with multiple cores that may have multiple vector processing units (VPUs).

The focus of this work is on exploiting multi-level parallelism through scheduling~\cite{blellochMultiscale}. 
Modern (and future) HPC systems (will) exhibit massive parallelism at different hardware levels.
Each of these hardware parallelism levels has its own corresponding scheduling technique that manages and schedules its computational load. 
For instance, batch, application, and thread level scheduling exist at cluster, node, and core levels, respectively. 
It is important to examine the relation between the different scheduling levels to take full advantage of the different hardware and software parallelism levels of modern HPC systems. 
For instance, making the individual schedulers at the batch and application levels aware of each other's decisions enables them to work \emph{in concert} for an optimized execution of applications and improved utilization of the underlying resources.

Simulation-based approaches are widely used in the literature for studying various aspects of single level scheduling. 
Yet, to the best of our knowledge, no simulators exist that support the study of two levels of scheduling. 
The main contributions of this work are: 
(1)~Introduces a generic simulation approach by bridging two existing simulators from batch and application scheduling levels;
(2)~Develops and evaluates a new \mbox{two-level} scheduling simulator based on GridSim~\cite{buyya2002gridsim} and SimGrid~\cite{casanova2014versatile}; and
(3)~The novel \mbox{two-level} simulator generates standard OTF2~\cite{eschweiler2011open} traces that allows visualization of the execution of jobs and tasks on allocated resources; 

In this work, combinations of three BLS algorithms, FCFS, EDF, and SJF, and four ALS algorithms, STATIC, SS, GSS, and FAC, are performed to explore the relation between BLS and ALS during operation. 
These combinations are used to schedule two real batch workloads from HPC production systems. 
Certain customizations are applied to these batch workloads to envision the missing information that is  necessary to simulate ALS.

The paper is structured as follows: 
In Section~\ref{sec:rewo}, the most relevant work in the literature is discussed. 
The background required to understand the BLS and ALS algorithms which are selected for the present work is reviewed in Section~\ref{sec:background}. 
In Section~\ref{subsec:appr}, a generic simulation approach for connecting two simulators of different scheduling levels is introduced. 
Based on this approach, in Section~\ref{subsec:two}, a new \mbox{two-level} simulator for BLS and ALS is defined.  
In Section~\ref{sec:exp}, the results of running real workloads with several combinations of BLS and ALS algorithms using the proposed \mbox{two-level} simulator are discussed.  
The information needed to reproduce this work is presented in Section~\ref{sec:reprod}. 
Finally, in Section~\ref{sec:conc}, the conclusions of this work and its potential future work are outlined.
\section{Related Work}
\label{sec:rewo}

Implementing, comparing, verifying, and validating a scheduling solution for HPC production systems involves numerous technical details and steps.
Therefore, simulation approaches have been widely used to examine the performance of different scheduling algorithms. 
Consequently, the crux of this section is to highlight the most relevant and influential simulators used for simulating scheduling of problems at the levels considered herein: batch and application scheduling, respectively.

\textbf{Simulation of BLS algorithms:}
One popular simulation toolkit is GridSim~\cite{buyya2002gridsim}. 
It facilitates simulation of grids, clusters, and single processing elements. 
It offers support for a broad range of heterogeneous resources including shared and distributed memory architectures. 
GridSim is built on top of a reliable discrete event simulation library called SimJava~\cite{simjava1998}. 
The GridSim toolkit is fully implemented in Java which promotes its portability and extensibility.

\textbf{Simulation of ALS algorithms:}   
SimGrid~\cite{casanova2014versatile} is a widely used simulation toolkit for ALS. 
Unlike GridSim (above) that focuses on simulating scheduling algorithms and resource allocation policies at grid and cluster levels, SimGrid supports the development of parallel and distributed applications in heterogeneous/homogeneous parallel and distributed environments. 
Recent releases of SimGrid have three different interfaces called: MetaSimGrid~(MSG), SimDag~(SD), and Simulated~MPI~(SMPI). 
The MSG interface allows users to define their applications as a group of concurrent processes, while the SD interface helps users to define their applications as directed~acyclic~graphs~(DAGs) models, instead of concurrent processes. 
SMPI has a unique capability of running unmodified applications written using the message~passing~interface~(MPI) in either simulation or real modes, and, thus, promotes the rapid development and testing of MPI applications. 
SimGrid can be considered a suitable candidate for a ALS simulation.

Due to the reliability and the active support community of the two toolkits, GridSim and SimGrid, they are used to build a large number of notable simulators, such as Alea~\cite{klusavcek2010alea} and Batsim~\cite{dutot2016batsim}.
Alea and Batsim are the most relevant simulators to the current work, and are used for the BLS and ALS simulations, respectively. 

Alea~\cite{klusavcek2010alea}~\cite{klusavcek2007alea} is a well-known simulator, developed on top of the GridSim toolkit. 
It extends GridSim and improves certain of its limitations. 
For instance, it supports reading of jobs files written in the standard workload format (SWF)~\cite{swfchapin1999benchmarks} and reading of platform files instead of defining the simulated platform within the code. 
Alea provides a group of data structures that enhance job and resource modeling. 
It also implements a set of scheduling algorithms and gives the opportunity to integrate other scheduling algorithms. 
It can be considered a suitable candidate for BLS simulations.
In the present work, simulations at the BLS level are performed with Alea.

Batsim~\cite{dutot2016batsim} is one of the most recent SimGrid-based simulators. 
It is based on the separation of concerns between system simulation and scheduling algorithms using two main components: \emph{batsim main} and \emph{batsim decision}. 
The main component is responsible for simulating the computational resources and it uses the SimGrid simulation toolkit underneath. 
The decision component is responsible for the scheduling decisions at the resources management level and it can be implemented in any programming language. 
Batsim depends on a Unix socket layer to allow communication between its two components.
An approach similar to the Batsim communication approach that depends on Unix sockets is used in this present work.

The GridSim and SimGrid toolkits are preferably used (not restricted) to support batch and application level scheduling, respectively. 
Certain research efforts are described below that studied extensions of one of these two simulation toolkits to support the simulation of other scheduling levels.\\
\textbf{ALS simulation based on a BLS simulation toolkit:}
In~\cite{srivastava2011enhancing}, the authors prove the ability of extending Alea to support ALS algorithms. 
The authors extended Alea to support ALS algorithms, such as Fixed~Size~Chunk~(FSC)~\cite{kruskal1985allocating}, static~chunking~(STATIC), self~scheduling~(SS)~\cite{tang1986processor}, guided~self~scheduling~(GSS)~\cite{polychronopoulos1987guided}, and factoring ~(FAC)~\cite{hummel1992factoring}. 
The work in~\cite{srivastava2011enhancing} carried over all the Alea's advantages to the ALS domain, such as application tasks being expressed in the SWF format and the effect of system failures being examined with different ALS techniques. 
However, the simulator provided by~\cite{srivastava2011enhancing} supports ALS algorithms in such a way that it can no longer support BLS algorithms.\\ 
\textbf{BLS simulation based on an ALS simulation toolkit:} 
Simbatch~\cite{caniou2008simbatch} is a \mbox{SimGrid-based} simulator. 
Simbatch uses the MSG interface of SimGrid to support simulations and development of BLS algorithms. 
Simbatch's uniqueness comes from the fact that it swaps the focus of SimGrid from the ALS perspective to the BLS perspective. 

All aforementioned simulators and simulation toolkits are designed to support \emph{single level} scheduling simulations, such as at the BLS or ALS. 
However, to explore the relation between multiple levels of scheduling, simulators are needed that can combine the required methods, tools, and techniques from the single-levels.
In~\cite{ciorba2012combined}, the concept of combining resource allocation (RA) with dynamic loop scheduling (DLS) was first proposed, under the name of ``CDS'', a combined dual-stage RA and DLS approach. 
CDS is a \mbox{two-stage} framework that maximizes the probability that applications complete by common deadline under certain levels of variation in the resources availability to compute and variation in system input. 
The RA techniques used in~\cite{ciorba2012combined} were initially simplistic and, subsequently, more sophisticated in~\cite{hansen2014heuristics}.


As a preliminary step for the work in the present paper, the original Alea simulator~\cite{klusavcek2010alea} has been \emph{redesigned} and \emph{reimplemented} to support ALS algorithms in addition to BLS algorithms, in~\cite{AhmedSC16}. 
Moreover, in~\cite{AhmedSC16} a new simulator based on the \mbox{SimGrid-SD}~interface~\cite{casanova2014versatile} was designed to support BLS algorithms in addition to ALS algorithms. 
These two simulators showed similar results in terms of total execution time for the simulated batches and applications at BLS and ALS, respectively. 
In the case of large batch workloads, the extended Alea simulator showed improved performance in terms of total simulation time, while the \mbox{SimGrid-SD-based} simulator showed improved performance in the case of applications that contain large numbers of tasks.  
The simulators presented in~\cite{AhmedSC16} support only one level of scheduling at a time: either BLS or ALS.

Attempting to simultaneously simulate the two levels of scheduling using the simulators in~\cite{AhmedSC16} revealed certain technical challenges.
In particular, both simulators are based on discrete events. 
Each maintains an individual simulation clock, updated according to the events occurring at the scheduling level they simulate. 
Extending any of the two simulators to support multiple simulation clocks involves numerous changes that may affect the functionality of simulation toolkit used to build that simulator.     
Another challenge is related to the initialization functions \texttt{GridSim.init} and \texttt{SD\_init} of GridSim and \mbox{SimGrid-SD}, respectively.
These functions were designed to be called one time at the beginning of the simulation. 
Thus, multiple calls would cause simulation errors during \mbox{execution}. 
In certain cases, the \mbox{SimGrid-SD-based} simulator would require multiple calls to \texttt{SD\_init} to reinitialize the simulation environment, when the simulator is launched for a different application.

In the present work, a novel simulation approach is proposed to develop two-level simulators. 
The proposed approach overcomes the challenges encountered in~\cite{AhmedSC16} by bridging scheduling simulators from two levels, in such a way that each remains responsible for simulating a specific scheduling algorithm at a certain level of hardware and software parallelism. 
Bridging simulators according to the proposed approach aims to minimize the changes in the single-level simulation toolkits.
Minimal changes in the simulation toolkits constitutes an advantage for obtaining the support of the user community and to naturally maintain compatibility with the new toolkit releases.


 \section{Background}
 \label{sec:background}
In this section, background information is presented to facilitate understanding of the BLS and ALS algorithms considered herein as well as details of the scheduling algorithms and their selection. 
 
\textbf{BLS algorithms} are used to allocate resources to non-interactive applications grouped in batches, along two dimensions: processor space and execution time. 
A non-interactive application is also referred to as a batch job. 
A job requests a certain number of resources to execute on from the batch-level scheduler.
BLS algorithms are implemented in queuing systems that assign free resources to jobs at the beginning of the queue.  
Based on their characteristics, jobs are prioritized and sorted using different algorithms, such as FCFS, EDF, and SJF~\cite{azmi2011scheduling}. 
In case no free resources to satisfy the allocation request of the job with the highest priority, the queuing system waits until a sufficient number of resources are free before assigning this job the requested resources. 
FCFS, EDF, and SJF are the most well-known BLS algorithms.
FCFS assigns the highest priority to the job with the minimum arrival time. 
EDF assigns the highest priority to the job with the minimum deadline.
SJF assigns the highest priority to  the job with the minimum execution time.
There are publicly available implementations of the FCFS, EDF, and SJF in the literature~\cite{klusavcek2010alea}. 
Therefore, they have been selected for the present work.
Other BLS algorithms will be considered in future work.

\textbf{ALS algorithms} are used to schedule, in space and time, the tasks within each application on parallel resources that are assigned to the application (during BLS) for a certain amount of time.
The performance of ALS for a given application depends on the algorithm and its implementation within the application, defined by the application developers. 
In HPC, a large number of scientific applications share a common characteristic: they consist of loops with large numbers of iterations.
Thus, loops are considered the main source of parallelism within HPC applications. 
For this reason, loop scheduling algorithms form an important subset of the ALS algorithms. 
In dynamic loop scheduling (DLS), the loop iterations are divided and distributed during the application 
\mbox{execution time} across all available processing elements.

A number of dynamic loop scheduling (DLS) are selected and considered in the present work: STATIC, SS~\cite{tang1986processor}, GSS~\cite{polychronopoulos1987guided}, and FAC~\cite{hummel1992factoring}. 
STATIC scheduling is also referred to as straightforward parallelization, wherein the loop iterations are divided into a fixed number of chunks equal to the number of available processing elements.
These chunks are assigned in a single round to the processing elements. 
Therefore, the scheduling overhead associated with STATIC is minimal. 
Yet, due to variability during execution time, severe load imbalance may arise. 
Alternatively, using SS, whenever a processing element is free, an individual iteration is assigned to it. 
The~loop iterations are assigned individually to the free processing elements. 
Thus, the use of SS incurs the largest scheduling overhead, resulting, however, in the most load balanced execution. 
Between these two extremes STATIC and SS, other two important DLS techniques have been proposed: GSS and FAC.
In GSS, the processing element executing the scheduler divides the total number of loop iterations into chunks of variable sizes, by dividing the remaining number of (unexecuted) loop iterations to the total number of processing elements.
Depending on the processing speeds, GSS may oversubscribe the first requesting processing elements with large chunks, the remaining loop iterations being insufficient to balance the rest execution on the processing elements.
The FAC technique overcomes this limitation of GSS by partitioning the number of loop iterations in batches of chunks. 
A batch consists of half of the remaining number of (unexecuted) loop iterations.
Each processing element is assigned a chunk of equal size from the batch. 
The present work focuses on the four aforementioned algorithms, as they represent important and widely known DLS algorithms. 
\section{The Proposed  Two-level Simulation Approach}
\label{sec:pro}
As discussed in Section~\ref{sec:rewo}, numerous simulators exist for scheduling in parallel and distributed systems. 
Each simulator has its own capabilities and limitations. 
It becomes necessary to define a set of objectives for the proposed two-level simulation approach, to crystallize its capabilities and limitations.\\
\textbf{Objectives of the proposed approach} \\
(\rom{1})~Preserve the current level of user involvement, to avoid that users learn new APIs or new simulation toolkits to perform their simulations.\\
(\rom{2})~The scheduling algorithms in the literature, at either of the BLS or ALS levels, can be easily `plugged' into the new simulator.\\
(\rom{3})~The approach exploits parallel computing systems with shared and/or distributed memory to reduce the overall simulation run-time.
\subsection {Bridging simulators via a connection layer}
\label{subsec:appr}
\begin{figure}
	\begin{center}
		\includegraphics[width=\columnwidth, clip, trim= 0cm 0cm 0cm 0cm]{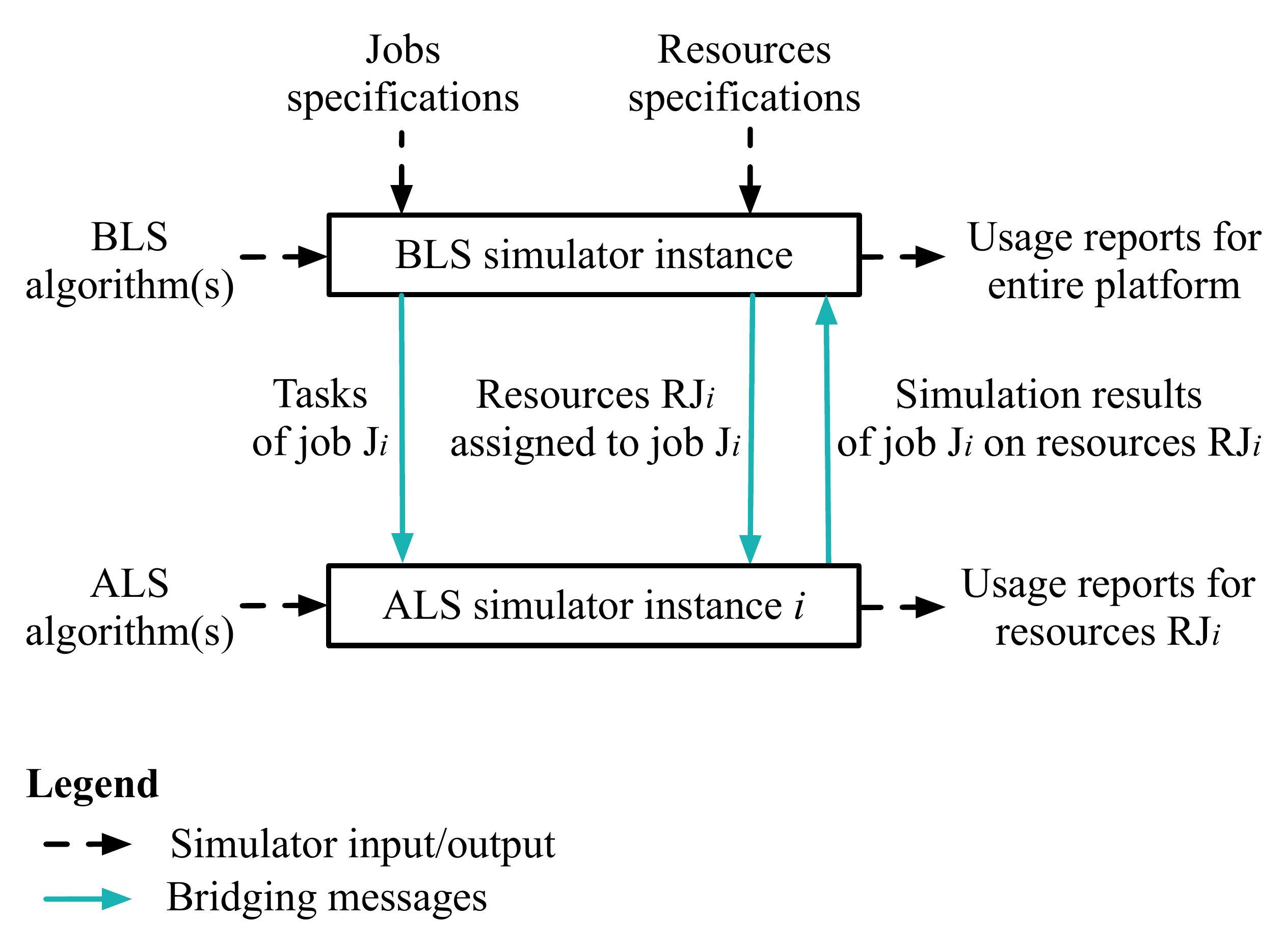}
		\caption{Bridging between simulator instances for allocating resources RJ$_i$ to job J$_i$ using a certain BLS algorithm and executing J$_i$ on RJ$_i$ according to a given ALS algorithm.}
		\label{fig:mainidea}
	\end{center}
\end{figure}

As shown in~\cite{AhmedSC16}, different simulators for parallel and distributed systems support simulation of scheduling algorithms at different levels of scheduling, i.e., BLS and ALS. 
However, certain simulators have a strong potential to support simulation of scheduling algorithms at certain levels of hardware parallelism. 
For instance, at the grid or cluster level, GridSim has the capability to implement and simulate BLS algorithms, while SimGrid has advantages to implement and simulate ALS algorithms. 
The two-level scheduling idea proposed in this work is, therefore, based on simultaneously executing two simulators such that each simulates a certain level of scheduling, both simulators feeding each other with their scheduling decisions when needed throughout execution. 
Fig.~\ref{fig:mainidea} illustrates an example in which the BLS simulator simulates a batch of jobs and requires as input three important parameters: the \textit{set~and~characteristics~of~the batch jobs}, the \textit{set~and~characteristics~of~the~cluster~resources}, and the \textit{BLS~algorithm}. 
The BLS simulator decides which cluster resources should be allocated to execute a certain job from the batch at a certain time. 
This decision is fed into the ALS simulator, instantiated for the particular job, with the specifications of the three parameters: \textit{tasks~of~particular~job}, \textit{description~of~allocated~resources}, and the \textit{ALS~algorithm}. 
The arrows in green color denote the connection layer between the two levels of scheduling. 
The illustrative example in Fig.~\ref{fig:mainidea} is extended in Fig.~\ref{fig:challe} (described later in Section~\ref{subsec:two}), in which the BLS simulator is \mbox{Gridsim-Alea} and the ALS simulators are instances of {SimGrid-SD}.
 

\subsection {The two-level simulator}
\label{subsec:two}
\begin{table}
	\centering
	\captionsetup{font=footnotesize,labelsep=periodspace,singlelinecheck=false, justification=centering}
	\caption{Notation}
	\label{tab:sym}
	\begin{tabular}{l|l}
		\textbf{Symbol} & \textbf{Description}\\ 	\hline
		$N$ & \begin{tabular}[l]{@{}l@{}}Number of batch jobs \end{tabular} \\ 
		$M$ & \begin{tabular}[l]{@{}l@{}}Number of cluster resources \end{tabular} \\ 
		$J$      & \begin{tabular}[l]{@{}l@{}}Set of batch jobs\\ $\{J_i \mid 0 \leq i \textless N\}$ \end{tabular} \\ 
		$R$      & \begin{tabular}[l]{@{}l@{}}Set of cluster resources\\ $\{R_j \mid 0 \leq j \textless M \}$ \end{tabular}\\ 
		$RJ_i$ & \begin{tabular}[l]{@{}l@{}}Set of resources allocated to job $J_i$\\ $RJ_i \subseteq R, RJ_i \neq \emptyset, 0 \leq i \textless N$ \end{tabular}\\ 
		$AT$   & \begin{tabular}[l]{@{}l@{}}Set of jobs arrival times \\ $\{AT_i \mid 0 \leq i \textless N\}$ \end{tabular} \\ 
		$FT$   & \begin{tabular}[l]{@{}l@{}}Set of jobs finishing times \\ $\{FT_i \mid 0 \leq i \textless N\}$ \end{tabular} \\ 
		$ST$   & \begin{tabular}[l]{@{}l@{}}Set of jobs starting times \\ $\{ST_i \mid 0 \leq i \textless N\}$ \end{tabular} \\ 
		$LJ_i$  & \begin{tabular}[l]{@{}l@{}}Length of job $J_i$ (in GFLOP),\\ where $0 \leq i \textless N$ \end{tabular} \\ 
		$TJ_i$&  \begin{tabular}[l]{@{}l@{}}Set of all tasks belonging to job $J_i$,\\ where $0 \leq i \textless N$ \end{tabular} \\ 
		$LT_k$  & \begin{tabular}[l]{@{}l@{}}Length of task $T_k$ (in GFLOP) of job $J_i$,\\ where $0 \leq k \textless |TJ_k|$ and $0 \leq i \textless N$ \end{tabular} \\ 
		$\Upsilon$  & \begin{tabular}[l]{@{}l@{}} Task variation factor\\ $0 \leq \Upsilon \textless 1$  \end{tabular} \\ 
		makespan $b$  & \begin{tabular}[l]{@{}l@{}} Time to complete all jobs of a certain workload,\\ where each job has an equal number of tasks\\ $\max(FT) - \min(AT)$  $|$ $ \Upsilon=0, \forall J_i \in J$ \end{tabular} \\ 
		makespan $\Upsilon$  & \begin{tabular}[l]{@{}l@{}} Time to complete all jobs of a certain workload,\\ where the sizes of tasks within each job varies \\ according to $\Upsilon$\\ $ \max(FT) - \min(AT)$  $|$ 
			$0 \textless \Upsilon \textless 1, \forall J_i \in J $\end{tabular} \\	 
	\end{tabular}
\end{table}

The proposed approach meets objectives (\rom{1}) and (\rom{2}) stated earlier in this section, due to the fact that users are free to select any two simulators for the two levels of scheduling and the scheduling algorithms are easily `plugged' in.
This approach helps users to leverage their expertise on the use of certain simulation toolkits.  
The proposed approach depends on simultaneously running several simulation instances as separate processes.
Moreover, these instances can simultaneously run on shared and/or distributed memory systems.     
Thus, the proposed approach also meets objective~(\rom{3}) stated earlier. 
All the results obtained within this work are based on running a new proposed two-level simulator on  a multi-core processor with shared memory, described below. 

Based on the approach discussed in Section~\ref{subsec:appr}, a new two-level simulator is designed and implemented by connecting and integrating two different simulators. 
The \mbox{GridSim-based} simulator, Alea~\cite{klusavcek2010alea}, is used to simulate BLS algorithms, while the \mbox{SimGrid-SD-based} simulator~\cite{AhmedSC16} is used to simulate ALS algorithms. 
Although both simulation toolkits are well-known and have an active support community, connecting them has not yet been attempted and poses certain implementation challenges.

The first challenge is interfacing two different programming models: object-oriented and structured programming used for developing the \mbox{GridSim-Alea} (in Java) and the \mbox{SimGrid-SD} (in C) simulators, respectively.
The second challenge is synchronizing the independent simulation clocks of the simulators instances. 
Both simulation toolkits are based on discrete events, and each keeps its own discrete simulation clock that is only advanced when an internal event occurs. 
The third challenge is merging the output results generated by the multiple instances of the two simulators to enable a proper informative presentation.

To address these challenges, a \emph{connection layer} (as a part of the proposed \mbox{two-level} simulator) was designed to provide the following functions: (i)~Manage simulator instances, (ii)~Synchronize the clocks of the simulator instances, and (iii)~Exchange necessary information regarding jobs, tasks, and other execution parameters.

To illustrate the use of the connection layer, consider the following scheduling scenario: 
A  batch $J$ consists of four jobs \mbox{$\{ J_0,J_1,J_2,J_3\}$}. 
Each job consists of three tasks.  
In each job, the sum of the lengths of the first two tasks is equal to the length of the third task, i.e., \mbox{$LT_1 + LT_2 = LT_3$.}
A cluster $R$ consists of five homogeneous resources \mbox{$\{R_0,R_1,R_2,R_3,R_4\}$}. 
The set of resources required by job $J_i$ is denoted $RJ_i$, $0 \leq i \textless 4$.
The following resource assignments are requested: \mbox{$RJ_0 = \{R_0,R_1\}$}, \mbox{$RJ_1 = \{R_2,R_3\}$}, \mbox{$RJ_2 = \{R_2,R_4\}$}, and \mbox{$RJ_3 = \{R_0,R_4\}$}. 
The arrival time of job $J_i$ is $AT_i$, $0~\leq~i~\textless~4$, where \mbox{$AT_0$  = $AT_1$ = $0$} and \mbox{$AT_2 < AT_3$}.
The finishing time of job $J_i$ is $FT_i$, $0~\leq~i~\textless~4$, and \mbox{$FT_0$  = $FT_1$ $ > AT_3 > AT_2$}.
FCFS and STATIC are used as BLS and ALS, respectively.

Since $AT_0$  = $AT_1$ = 0, the connection layer \textbf{manages} the BLS and ALS simulator instances by \emph{starting} two separate instances of the \mbox{SimGrid-SD-based} simulator to simulate the execution of jobs $J_0$ and $J_1$ on $RJ_0$ and $RJ_1$, respectively, using STATIC. 
Given that \mbox{$FT_1 > AT_2$} and \mbox{$RJ_1 \cap RJ_2 = \{R_2\}$}, the connection layer \emph{holds} the simulation of $J_2$ until the \mbox{SimGrid-SD-based} simulation instance for $J_1$ reports its completion.  
Since \mbox{$AT_3 > AT_2$}, $J_2$ starts before $J_3$, and, thus, the connection layer holds the simulation of $J_3$ until the \mbox{SimGrid-SD-based} simulation instances for $J_0$ and $J_2$ report their completion, given that \mbox{$RJ_3 \cap RJ_0 = \{R_0\}$} and \mbox{$RJ_3 \cap RJ_2 = \{R_4\}$}. 
Therefore, the time at which simulation of $J_3$ begins depends on the times at which the simulation of $J_0$ and $J_2$ completes. 
The finishing times of $J_0$ and $J_2$ are dominated by the scheduling decisions of the ALS algorithms. 
Recall that for jobs $J_0$ and $J_2$, the sum of the lengths of the first two tasks equals the length of the third task.
Due to using STATIC as ALS and having homogeneous resources, load imbalance will arise in executing the three tasks of $J_0$ and $J_2$ on the sets of resources $RJ_0$ and $RJ_2$ that are assigned to $J_0$ and $J_2$. 
As a consequence, the BLS scheduler, FCFS, needs to delay the beginning of the execution of $J_3$.
The influence between BLS and ALS becomes visible via the fact that STATIC as ALS not only affects the individual performance of $J_0$ and $J_2$, but also the performance achieved by FCFS as BLS for scheduling the other jobs in the batch. 
In this scenario, if the FCFS algorithm passed certain information to the STATIC algorithm to prioritize the release of resources, the STATIC algorithm would assign the smallest chunk of tasks to the resources needed to be released for other jobs, such as $R_0$ and $R_4$.

\begin{figure}
	\includegraphics[width=\columnwidth, clip, trim= 0cm 0cm 0cm 0cm]{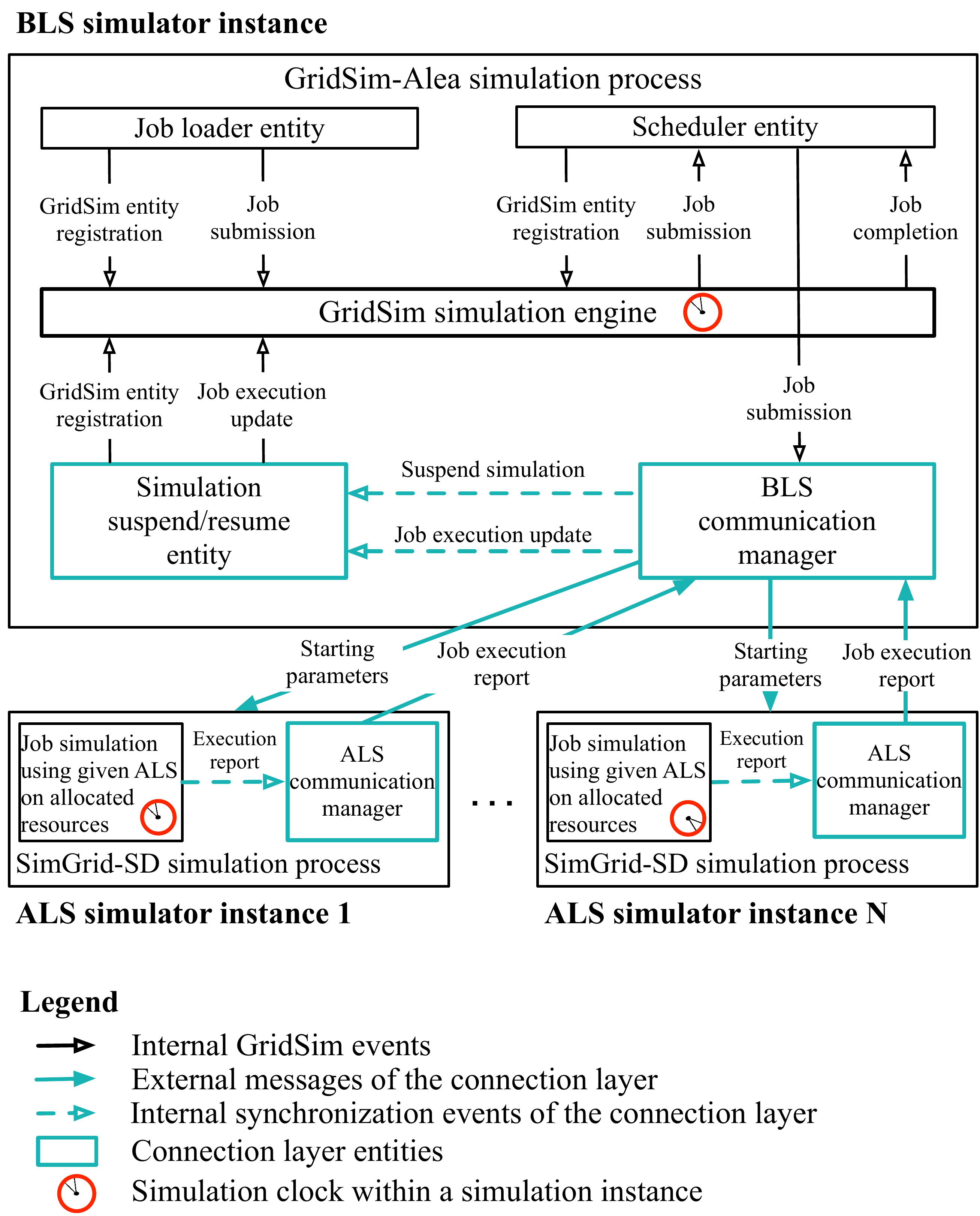}
	\caption{Two-level scheduling consisting of a single BLS and several ALS simulation instances. The connection layer synchronizes the independent simulation clocks of the \mbox{GridSim-Alea-based} and \mbox{SimGrid-SD-based} simulators.}
	\label{fig:challe}
\end{figure}

To support this type of scenarios, the connection layer \textbf{synchronizes} the running simulators using two strategies: \emph{simulation suspend/resume} and \emph{event injection}, as illustrated in Fig.~\ref{fig:challe}.
\mbox{A \textit{simulation suspend/resume entity}} registered in \mbox{GridSim-Alea}, is used to \emph{suspend} and \emph{resume} the BLS simulation. 
It performs a busy loop that ends \textit{if and only if} all running instances of the \mbox{SimGrid-SD-based} simulator report their completion and results. 
Due to the fact that the suspend entity is a registered GridSim entity, its busy loop can pause the simulation clock of the \mbox{GridSim-Alea-based} simulator until the busy loop ends. 

The internal synchronization events in Fig.~\ref{fig:challe} created by the BLS communication manager are used to update the simulation suspend/resume entity.
Thus, the \mbox{suspend/resume entity} can incrementally \emph{inject} the execution reports of the running \mbox{SimGrid-SD} simulation instances into the \mbox{GridSim} engine (see Fig. \ref{fig:challe}) and end its busy loop when there are no more running \mbox{SimGrid-SD} simulation instances.
The \mbox{simulation suspend/resume entity} injects the execution reports as \textit{GridSim events}. Therefore, the GridSim engine is able to use them to advance its simulation clock.
 Fig.~\ref{fig:challe} depicts the independent simulation clocks of the \mbox{GridSim-Alea-based} and \mbox{SimGrid-SD-based} simulators and their synchronization by connection layer.
The connection layer uses socket-based communication and application arguments to \textbf{exchange} the information between the \mbox{GridSim-Alea-based} simulator and the \mbox{SimGrid-SD-based} simulator instances.  
The connection layer launches \mbox{SimGrid-SD-based} simulator instances as independent application processes, and passes certain parameters as application arguments to each launched process. 
For example, it sends the port number on which it expects to receive the ALS simulation results.

\section{Experiments and Results}
\label{sec:exp}
To explore the relation between BLS and ALS algorithms in HPC systems and its impact on the batch and application execution, this work considers real workloads from production HPC systems. 
These workloads are selected from the \textit{parallel workload archive}~(PWA)~\cite{feitelson2014experience}~\cite{parallelarchive2016}. 
The PWA is a public archive containing detailed information on 38 workloads of large scale parallel machines from around the world, collected between 1993 and 2015. 
The workloads are provided in the standard workload format (SWF). 
A workload may exist in two versions: cleansed and raw. 
The cleansed version is a minor modification of the raw version to guarantee \mbox{self-consistency}. 
The founders of PWA recommend to use the cleansed version, when available. 
In this work, the cleansed versions of two different workloads have been used: CEA CURIE (first French Petascale machine) and LLNL Thunder, referred to as $W_1$ and $W_2$, respectively. 
Table \ref{tab:2} summarizes their characteristics.

In all experiments reported herein, a simulated platform that consists of four hosts is used. 
Each of the hosts has a processor that contains 64~cores. 
The maximum performance of one host is 3~TFLOP/s.
A fully connected network topology is used to connect the four hosts.
The network model used herein is an InfiniBand model with a link bandwidth and latency of 50~Gbps and 500~ns, respectively.

%
  
   \begin{table}
 	\centering
 	\captionsetup{font=footnotesize,labelsep=periodspace,singlelinecheck=false, justification=centering}
 	\caption{Characteristics of the workloads selected from the PWA }
 	\label{tab:2}
 	\begin{tabular}{lc|c}
 	\multicolumn{1}{l|}{\textbf{Workload}} & $\mathbf{W_1}$ & $\mathbf{W_2}$ \\ \hline
 		\multicolumn{1}{l|}{Provenance} & \begin{tabular}[c]{@{}c@{}}Curie supercomputer\\ operated by CEA\end{tabular} & \begin{tabular}[c]{@{}c@{}}Thunder Linux cluster\\ operated by LLNL\end{tabular} \\ \hline
 		\multicolumn{1}{l|}{Period of time} & Feb, 2011 -- Oct, 2012 & Jan, 2007 -- Jun, 2007 \\ \hline
 		\multicolumn{1}{l|}{Total number of jobs} & 312,000 & 121,000
	\end{tabular}
 \end{table}  
\subsection {Workload customization}
\label{subsec:customizatoin}

The workloads from the PWA only keep information relevant for batch level scheduling, such as \textit{job~ID}, \textit{submission~time}, \textit{wait~time}, \textit{allocated~resources} and \textit{user~ID}. 
For the purpose of this work, additional details regarding the application level are important and needed, such as the characteristics of the application each job represents, the number of parallel tasks within the application, the resource usage by each task, etc. 
These details are necessary for performing application level scheduling in the present work. 
Since this information is not present in the PWA workloads, certain assumptions are made about the applications:\\
(\rom{1})~All jobs in the workload are computationally-intensive. 
Consequently, all communication or I/O tasks that may exist in the original jobs are not considered in this work.
This assumption is not a limitation of the proposed approach. 
It is simply used to convert the existing workloads to one of the possible cases where jobs are computationally-intensive, as such jobs are among the main incentives for using HPC systems.
Other forms of jobs, such as \mbox{communication-intensive}, \mbox{I/O-intensive} and combinations thereof will be considered in future work.\\
(\rom{2})~Although the number of tasks and length of each task are application-dependent, this work considers the case of \emph{ideal parallelism}: all available hardware parallelism is exploited, execution is perfectly load balanced, communication is virtually instantaneous, and the resources allocated to tasks are identical. this work considers the case of ideal parallelism as an important baseline case.

In addition, other cases are generated and examined by introducing variation at the task length level using the task~variation~factor~$\Upsilon$.
By considering job $J_i$ and its allocated set of resources $RJ_i$, the elements of the set $TJ_i$ of tasks of job~$J_i$ can be randomly generated according to a probability distribution with mean~\mbox{$\mu = \dfrac{LJ_i}{|RJ_i|}$} and standard~deviation~\mbox{$\sigma = \mu \times \Upsilon$.}

Workloads from the PWA store the execution time and the number of resources requested by each job. The PWA also states certain detailed about the hardware platform where the workloads were obtained from. 
Based on his information, an estimate about each job length (in GFLOP) can be obtained by multiplying the minimum performance of request resources(in GFLOP/s) to the job execution time (in Second).
The job length $LJ_i$ is deduced by accumulating the length of the generated tasks, until $LJ_i$ become greater than or equal to the estimate.

Many researchers modeled the arrival, finishing, execution times, and number and type of requested resources of different jobs in the context of HPC~\cite{lublin2003workload,cirne2001model}. 
Few efforts exist that can be used to model (with certain adaptations) the number of tasks and the task length within HPC applications~\cite{pfeiffer2008modeling}.
In the present work, for simplicity, the task lengths are generated according to normal distribution with the aforementioned $\mu$ and $\sigma$ parameters. 
Future work will consider other models to generate the task lengths within jobs of a given workload.

\subsection {Analysis of the relation between BLS and ALS algorithms}

To explore  the relation between BLS and ALS algorithms, experiments were performed following two different strategies. 
The first strategy supports a \textit{coarse-grain analysis} of the relation between BLS and ALS, examining the effect of changing the ALS algorithms on the BLS performance, measured as the makespan of the entire batch workload. 
The second strategy supports a \textit{fine-grain analysis} of the relation between BLS and ALS that examines the effect of changing the ALS algorithm in one job on the starting time of its successor job(s) in the batch. 

In this work, combinations of three BLS algorithms FCFS, EDF, and SJF~\cite{azmi2011scheduling} and four ALS algorithms STATIC, SS~\cite{tang1986processor}, GSS~\cite{polychronopoulos1987guided}, and FAC~\cite{hummel1992factoring} described earlier in Section~\ref{sec:background}, are considered. 
To perform the experiments for the coarse-grain analysis, jobs of the most intensive 24 hours in terms of job arrival time have been selected from both $W_1$ and $W_2$, respectively. 
These most intensive \mbox{24-hour} intervals of $W_1$ and $W_2$ are referred to as $W^{24}_{1}$ and $W^{24}_{2}$, respectively.
A \mbox{Java-based} tool was developed as part of this work and used to extract the jobs of the most intensive period of 24 hours from a given workload.

   		
   \begin{figure}
   	\includegraphics[width=\textwidth, clip, trim= 0cm 6.5cm 16cm 1cm]{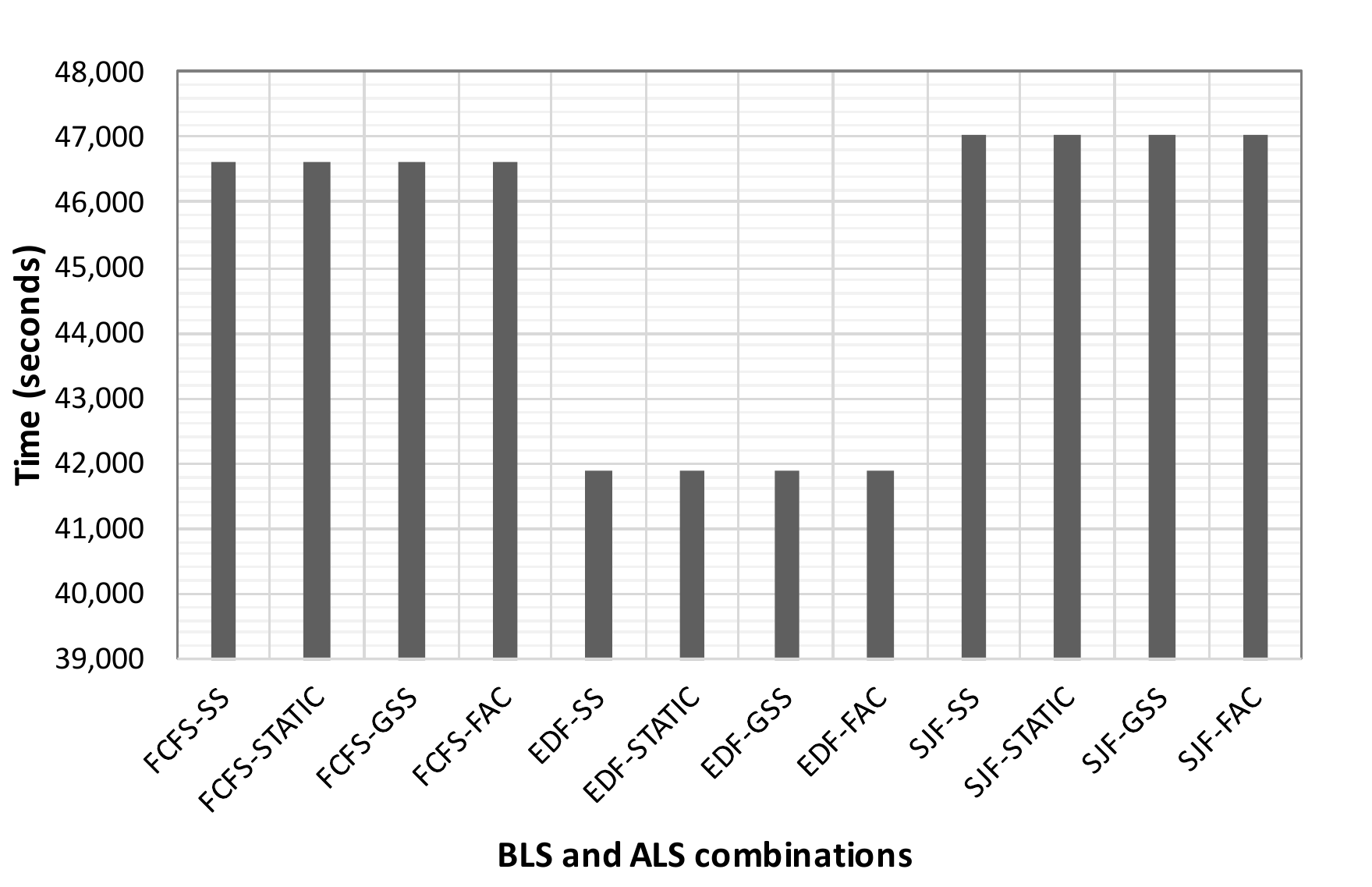}
   	\caption{The makespan of the $W^{24}_{1}$ workload that consists of 1,700 jobs for the twelve combinations of selected BLS and ALS algorithms}
   	\label{fig:w1_nox}
   \end{figure}
   \begin{figure}
   	\includegraphics[width=\textwidth, clip, trim= 0cm 6.5cm 16cm 1cm]{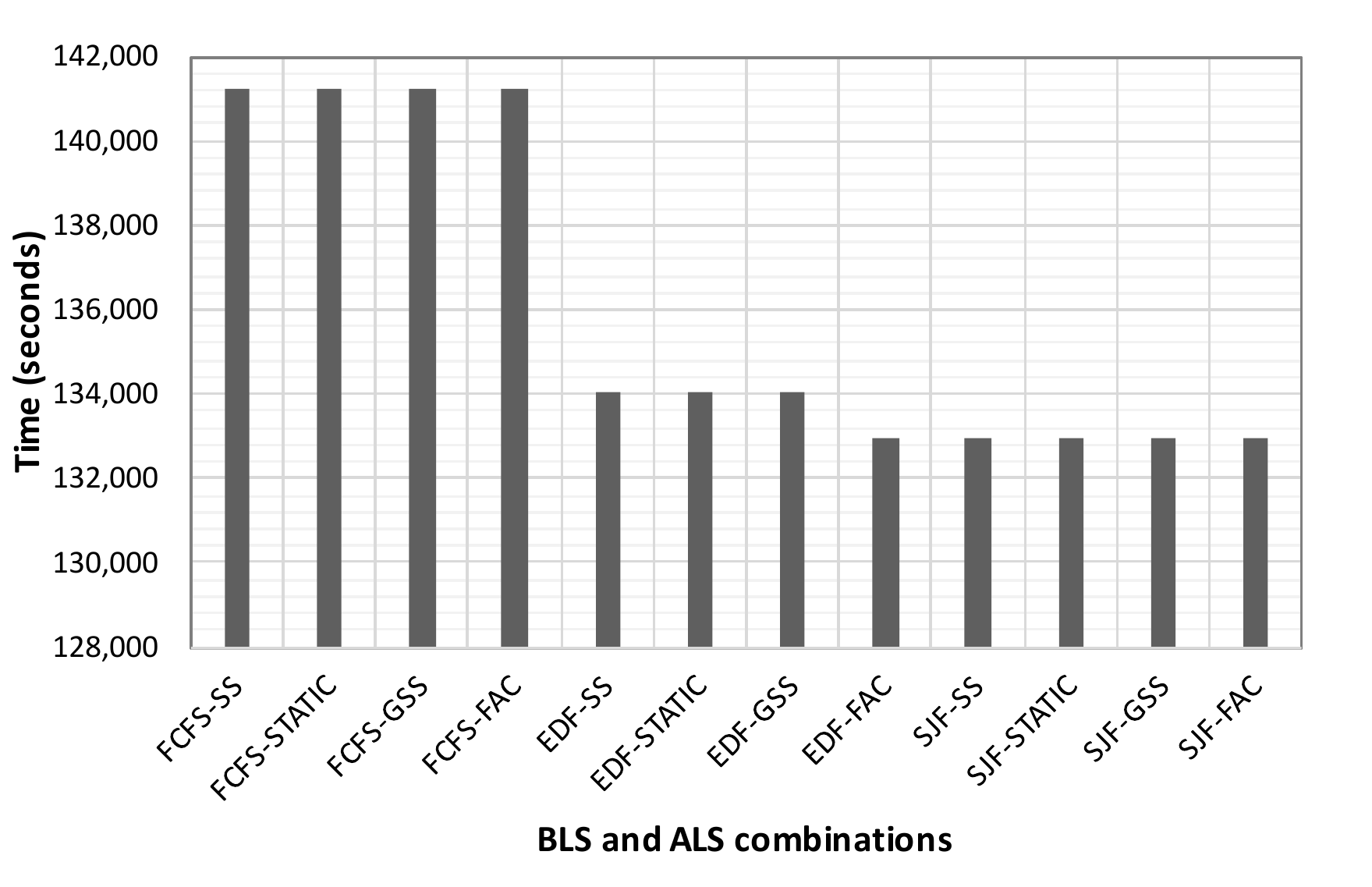}
   	\caption{The makespan of the $W^{24}_{2}$ workload that consists of 3,100 jobs for the twelve combinations of selected BLS and ALS algorithms}
   	\label{fig:w2_nox}
   \end{figure}
Fig.~\ref{fig:w1_nox} and Fig.~\ref{fig:w2_nox} show the total makespan of $W^{24}_{1}$ and $W^{24}_{2}$, respectively, when executing using the twelve combinations of the four BLS and the three ALS algorithms.
Each job in the two workloads is divided into a number of identical length tasks equal to the number of allocated resources. 
The task length variation factor $\Upsilon$ is not used in these experiments.
The results of Fig.~\ref{fig:w1_nox} and Fig.~\ref{fig:w2_nox} correspond to the best case scenario in which all submitted applications are perfectly optimized for their allocated resources. 
Although such a scenario is highly desirable both at the cluster operation level and at the user level, it is difficult to be encountered in practice.

\begin{figure}
	\includegraphics[width=\textwidth, clip, trim= 0cm 6.5cm 16cm .5cm]{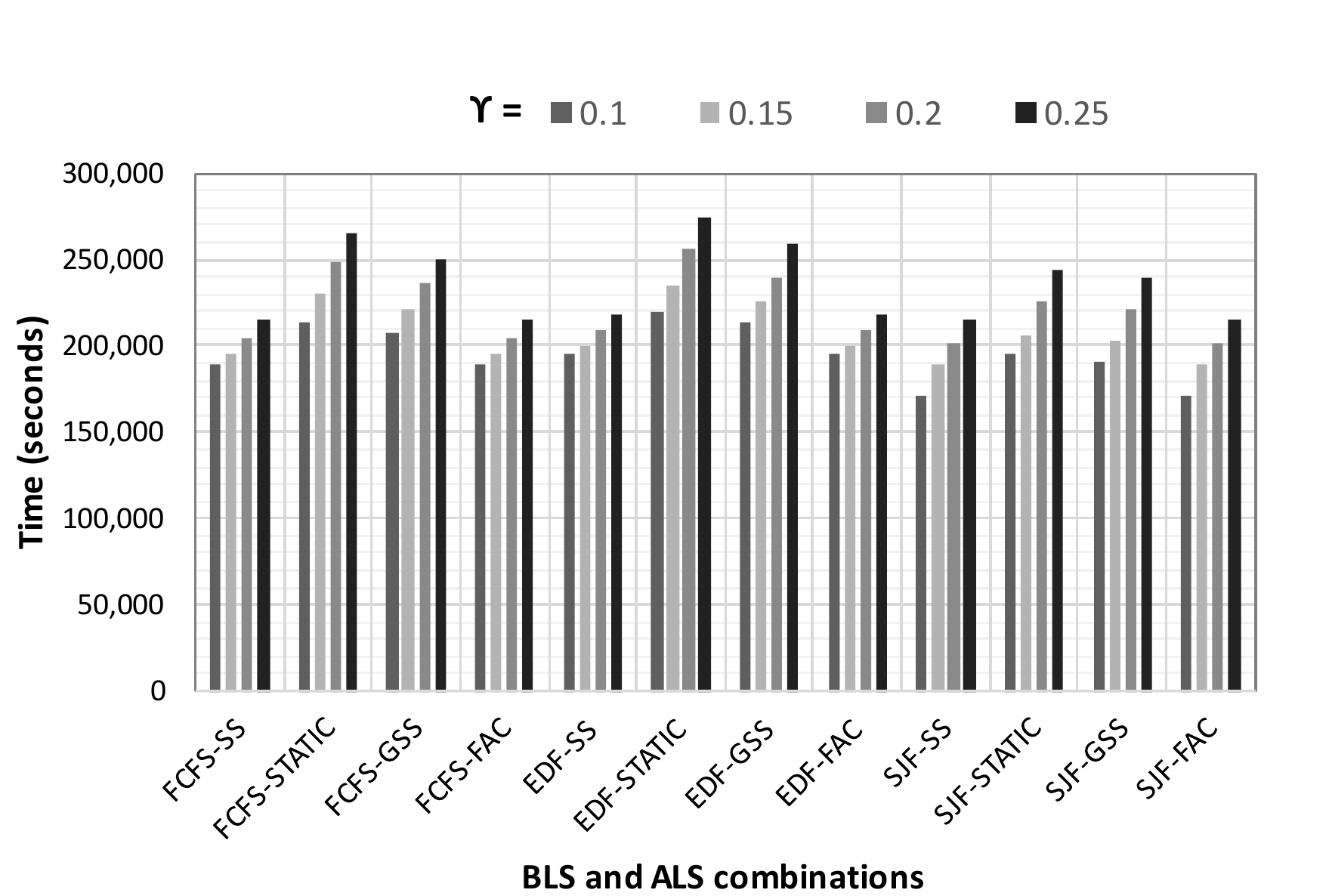}
	\caption{Effect of changing the task variation factor $\Upsilon$ from 0.1 to 0.25 on the total workload makespan for the twelve combinations of selected BLS and ALS algorithms for the jobs within $W^{24}_{1}$}
	\label{fig:w1_x}
\end{figure}

\begin{figure}
	\includegraphics[width=\textwidth, clip, trim= 0cm 6.5cm 16.5cm 1.5cm]{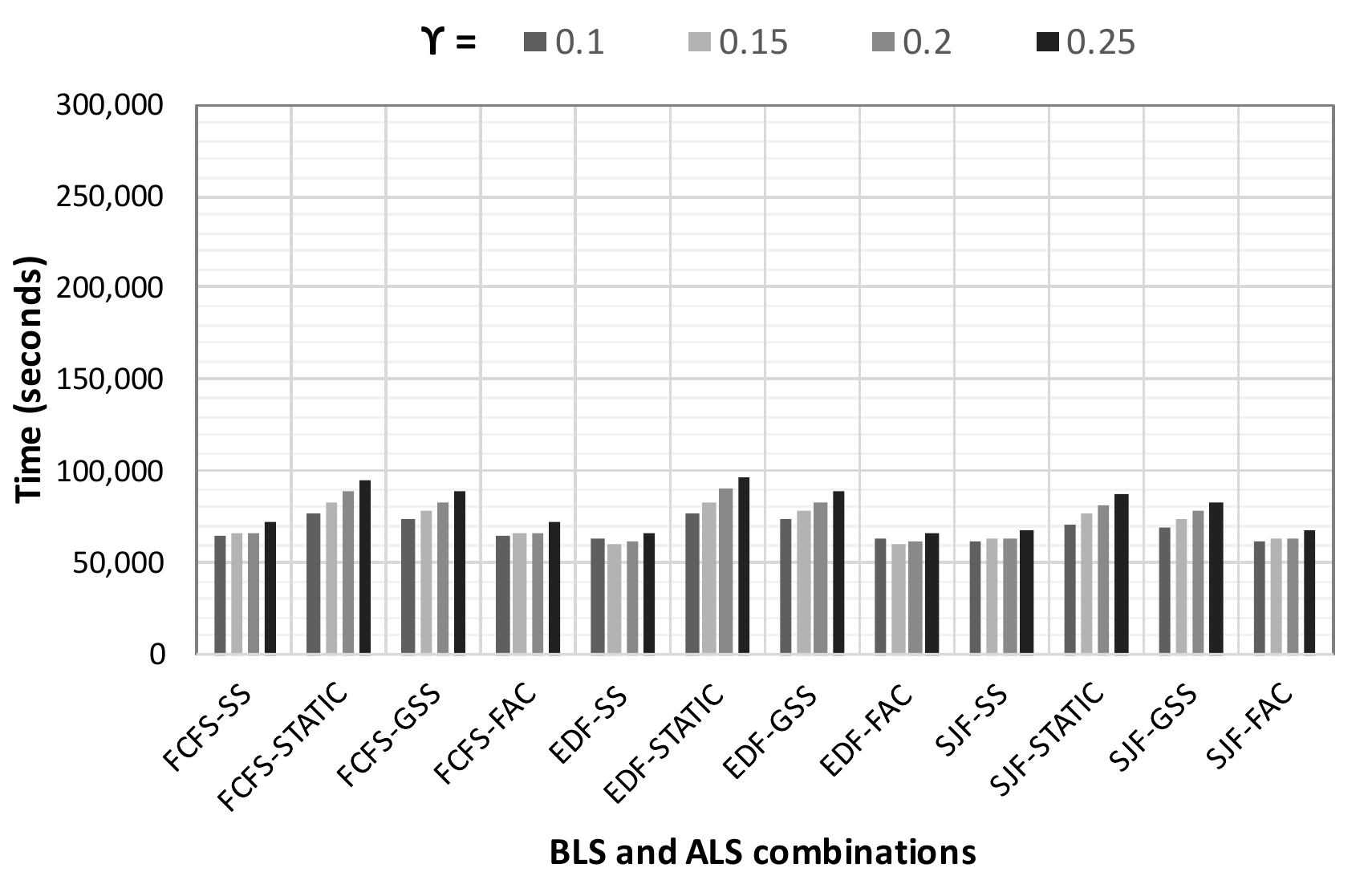}
	\caption{Effect of changing the task variation factor $\Upsilon$ from 0.1 to 0.25 on the total workload makespan for the twelve combinations of selected BLS and ALS algorithms for the jobs within  $W^{24}_{2}$}
	\label{fig:w2_x}
\end{figure}

As discussed in Section~\ref{subsec:customizatoin}, the task length variation factor $\Upsilon$ is used to  vary the lengths of tasks within certain job to represent more realistic applications. 
The results in Fig.~\ref{fig:w1_x} and Fig.~\ref{fig:w2_x} show the effect of increasing $\Upsilon$ from~0.0 (as considered in Fig.~\ref{fig:w1_nox} and Fig.~\ref{fig:w2_nox}) to 0.1, 0.15, 0.2, and 0.25, respectively for the twelve combinations of BLS and ALS algorithms.
From the results in Fig.~\ref{fig:w1_x} and Fig.~\ref{fig:w2_x}, one can infer that increasing $\Upsilon$ leads to an increase in the total makespan of both workloads~$W^{24}_{1}$~and~$W^{24}_{2}$, regardless of the BLS-ALS combination used. 
The amount of time corresponding to this increase in total makespan for different BLS-ALS combinations is not constant. 
Certain combinations showed the ability to better absorb the effect of increasing $\Upsilon$ than others. 
Further analysis is needed to understand this behavior and pinpoint its root-cause(s).
This type of insight can be used to enhance existing batch level schedulers and, consequently, two-level scheduling in which they are considered.

\begin{figure}
	\includegraphics[width=\textwidth, clip, trim= 0cm 7cm 16.5cm 1.8cm]{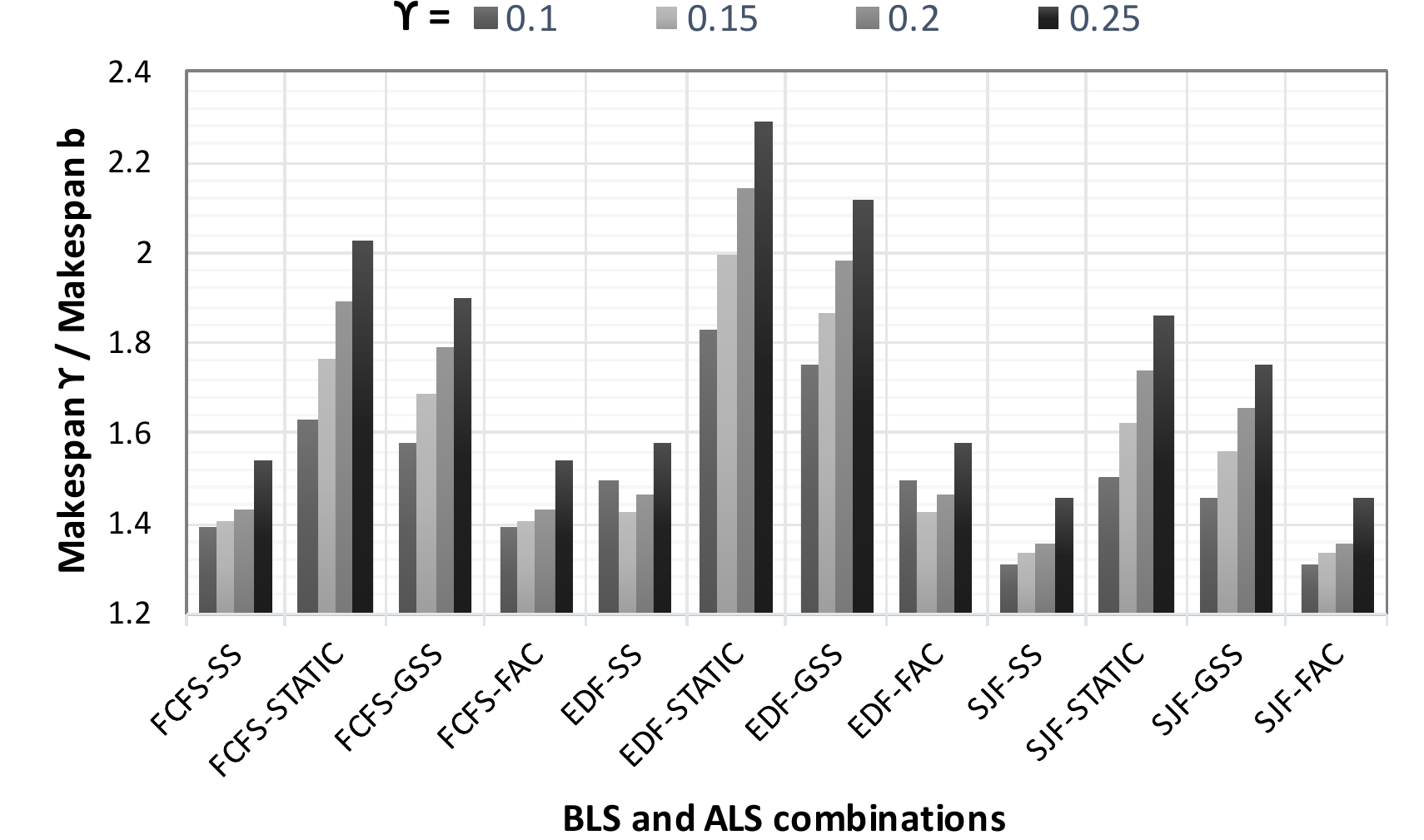}
	\caption{The ratio between makespan $\Upsilon$ and makespan $b$ for the twelve combinations of selected BLS and ALS algorithms for the jobs within $W^{24}_{1}$}
	\label{fig:rew1}
\end{figure}

\begin{figure}
	\includegraphics[width=\textwidth, clip, trim= 0cm 6cm 16cm 2cm]{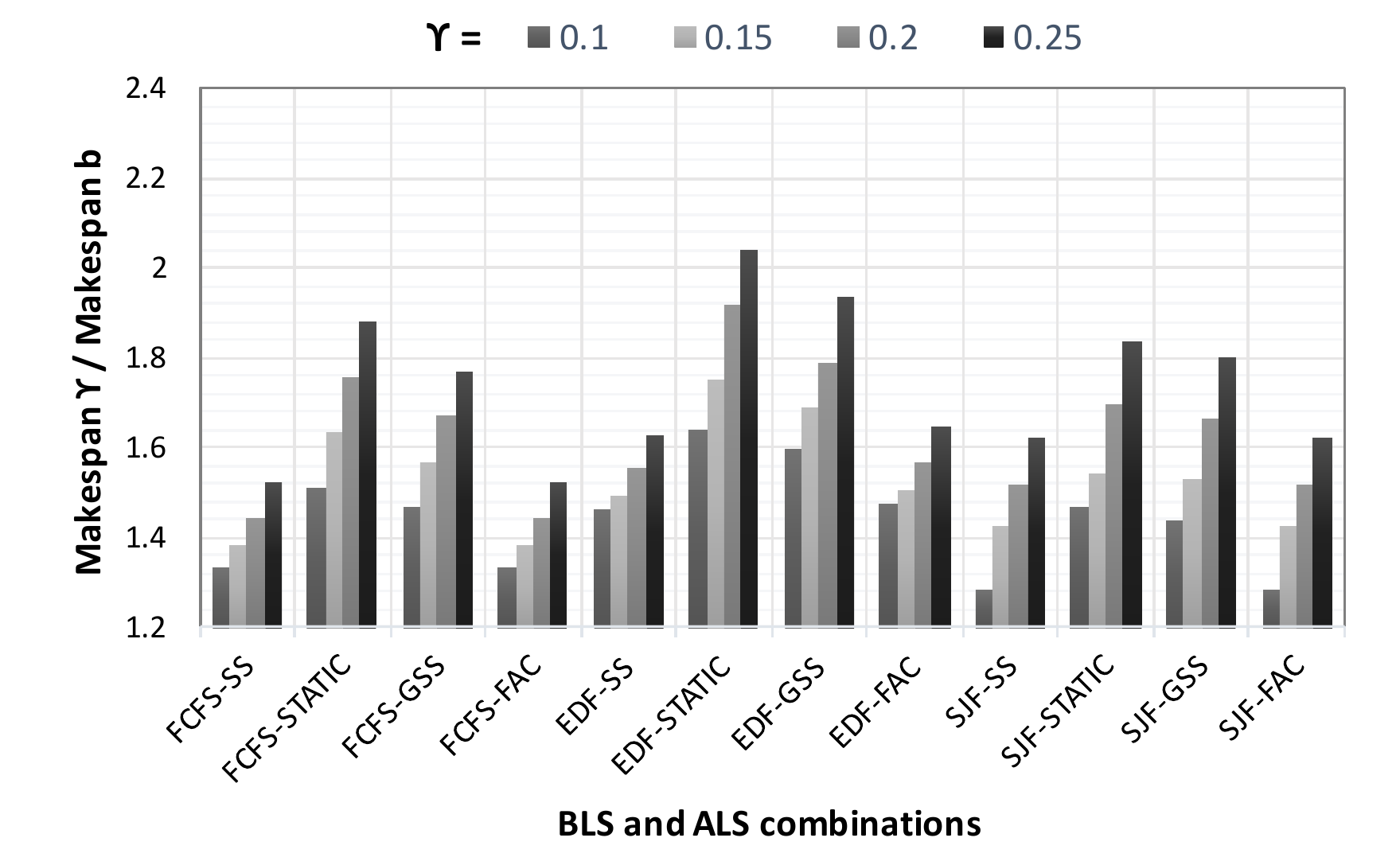}
	\caption{The ratio between makespan $\Upsilon$ and makespan $b$ for the twelve combinations of selected BLS and ALS for the jobs within $W^{24}_{2}$}
	\label{fig:rew2}
\end{figure}

The results in Fig.~\ref{fig:rew1} and Fig.~\ref{fig:rew2} show the ratio between two important performance measurements: makespan~$\Upsilon$ and makespan~$b$.
Makespan~$\Upsilon$ and makespan~$b$ are the total amounts of time required to complete all jobs of a certain batch of jobs in the presence and absence of $\Upsilon$, respectively. 
The ratio $\dfrac{ \text{makespan}~\Upsilon}{\text{makespan}~b}$ can be used to characterize the immunity of the system performance to given BLS-ALS combinations.

To perform a fine-grain analysis, the connection layer between \mbox{GridSim-Alea-based} and \mbox{SimGrid-SD-based} simulators was extended with an additional task: to collect all text-based traces generated from the \mbox{SimGrid-SD-based} simulator and to combine them into a single text-based trace file. 
The main challenge associated with this task is that each instance of the \mbox{SimGrid-SD-based} simulator does not have the global view of the entire batch workload simulation. 
For instance, to simulate jobs $J_1$ and $J_2$ on the sets of resources $JR_1$ and $JR_2$ at times $t_1$ and $t_2$, respectively, the connection layer runs two instances of the \mbox{SimGrid-SD-based} simulator. 
Each SimGrid instance, however, simulates its corresponding job as $J_i$ on the set of resource $JR_i$ at time $t_x$.

\begin{figure*}
	\includegraphics[width=\textwidth, clip, trim= 0cm 3.5cm 0cm 2cm]{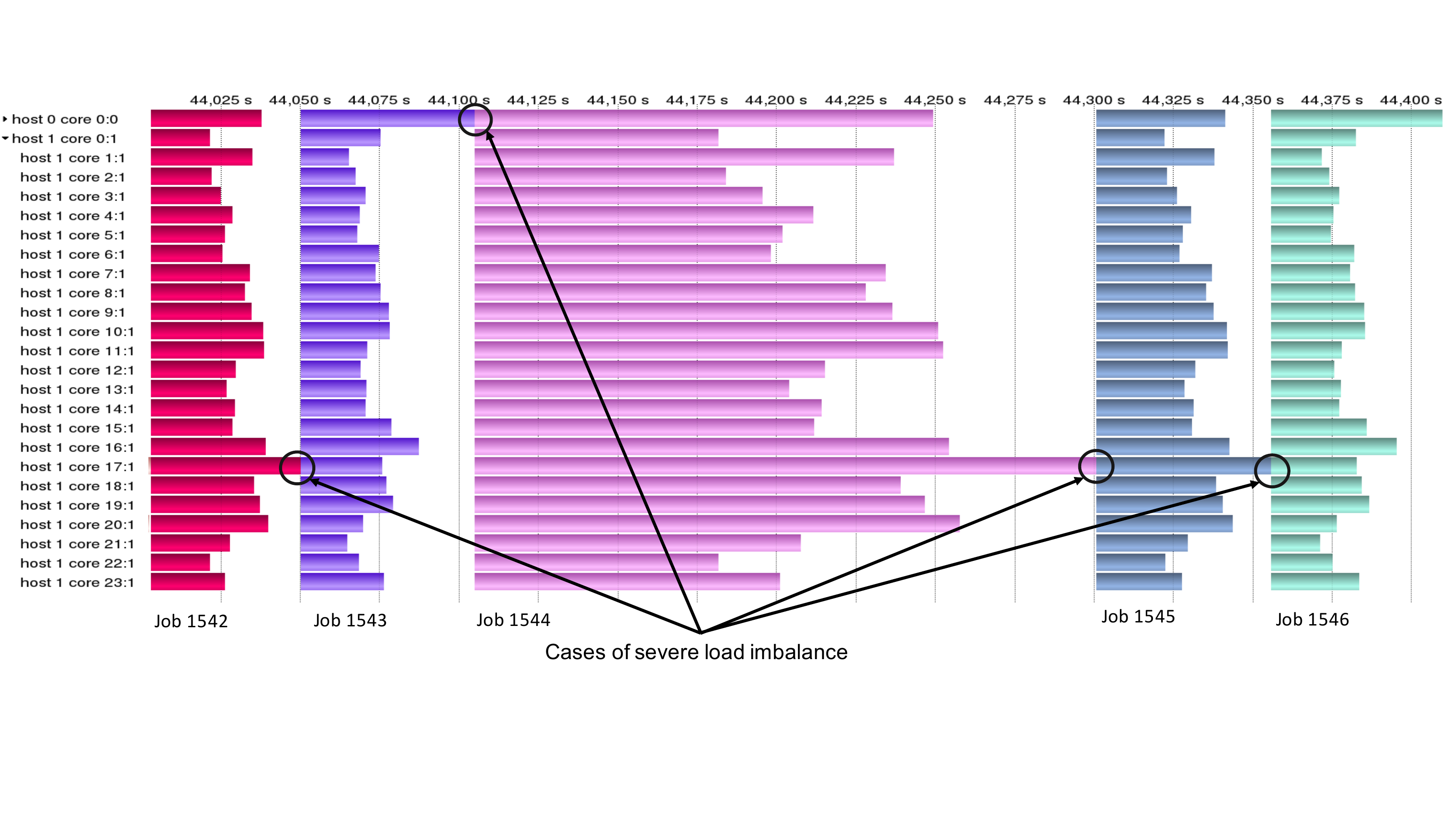}
	\caption{Snapshot of the Vampir  visualization tool showing the generated OTF2 trace of the proposed two-level scheduling simulator. The execution of different jobs and their tasks are shown according to their allocated resources at node and core levels, respectively. Tasks of the same job are represented using horizontal bars of the same color, while the white space between the job bars represents the idle state of the allocated cores. For simplicity, this snapshot shows five different jobs running over four simulated nodes (hosts). The illustration only contains 24 cores of host 1, while host 0 is collapsed, and hosts 2 and 3 are not shown. The scheduling algorithms shown herein are FCFS and
		GSS at BLS and ALS, respectively. Jobs are obtained from workload $W_1^{24}$}
	\label{fig:gview}
\end{figure*}

In this work, a tool was used to convert the collected text-based traces to binary traces in the OTF2~\cite{eschweiler2011open} format. 
Using OTF2 traces with the Vampir~\cite{knupfer2008vampir} trace visualizer, we are able to visualize for the first time, to the best of our knowledge, the cluster utilization from the node to the core level \textit{and} from batch level to application level scheduling, as shown in Fig.~\ref{fig:gview}. 
A snapshot captured from Vampir is included in Fig.~\ref{fig:gview} and shows the execution of five out of 1,700 running jobs, namely $J_{1542}$, $J_{1543}$, $J_{1544}$, $ J_{1545}$, and $J_{1546}$, from the $W^{24}_{1}$ workload.
The execution was performed with a combination between FCFS as BLS and GSS as ALS. 
The tasks of the five different jobs utilize host $1$. 
The other three hosts are also utilized by the five jobs.
Due to limited space, the execution of the five jobs on hosts $0$, $1$, and $2$ are collapsed and not shown.
Fig.~\ref{fig:gview} illustrates a case of severe load imbalance of certain jobs, its effects on the starting times of subsequent jobs in the batch, and, consequently, the effects on the entire system performance and utilization.

\begin{figure}
	\includegraphics[width=\textwidth, clip, trim= 0cm 6.2cm 17cm 0cm]{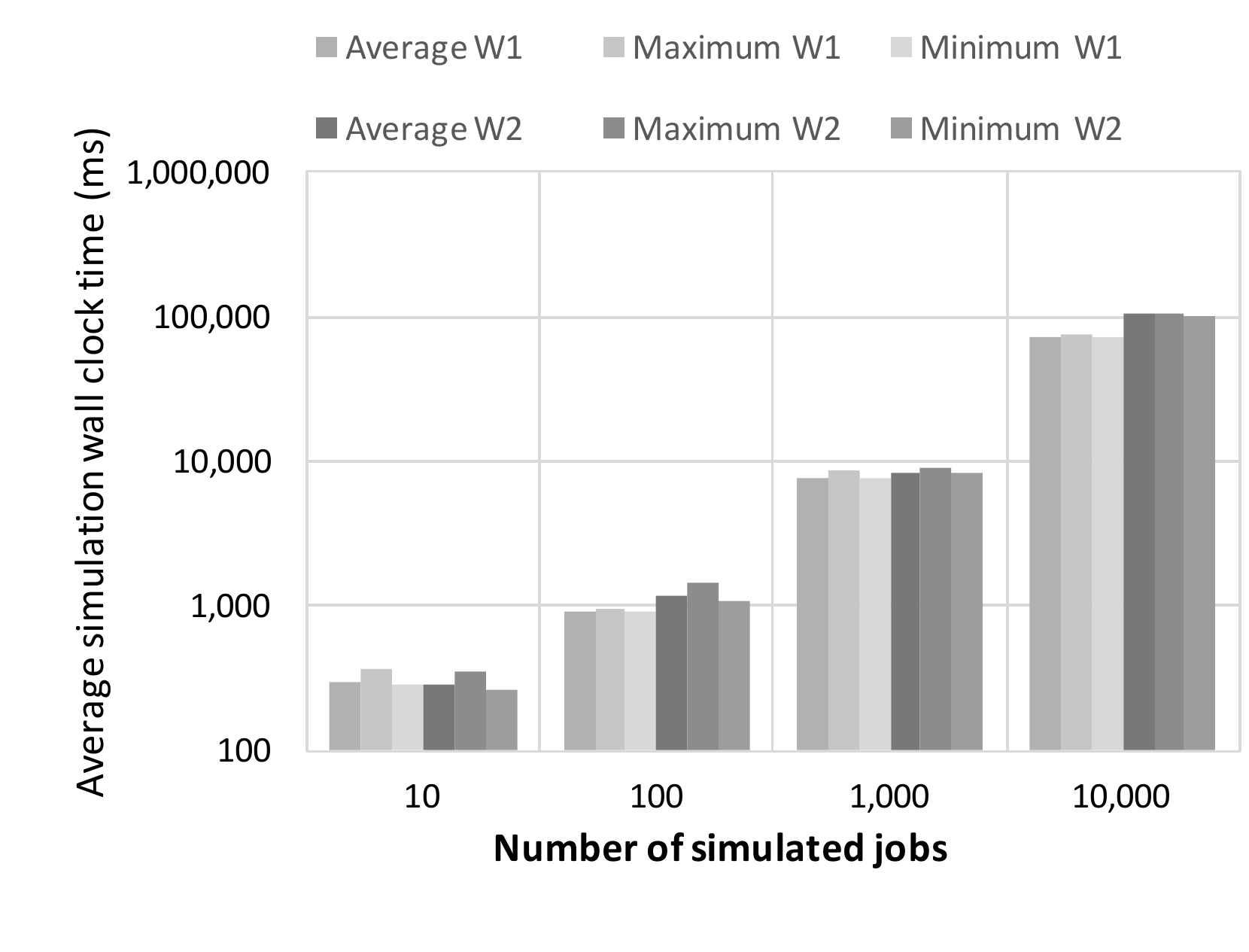}
	\caption{The simulation wall clock time (minimum, average, maximum) of the \mbox{two-level} simulator on an increasing number of jobs from workloads~$W_1$~and~$W_2$}
	\label{fig:allsimulation-clock}
\end{figure}

Scalability is an interesting aspect of the proposed two-level simulation approach, in terms of increasing the number of jobs and, consequently, in terms of increased number of simultaneous ALS instances.
An initial scalability assessment of the \mbox{two-level} simulator is presented in
Fig.~\ref{fig:allsimulation-clock}.
In these experiments, the average of the simulation wall clock times are reported for executing an increasing number of jobs (from 10 to 10,000) from the two workloads $W_1$ and $W_2$ described earlier in Table~\ref{tab:2}.
The simulation wall clock is defined as the total time required by the \mbox{two-level} simulator to simulate the execution of all jobs of a given workload and their tasks. 
The experiments were conducted with the least performing BLS-ALS combination, i.e., FCFS-STATIC, with $\Upsilon=0.25$. selected from the results in Fig.~\ref{fig:w1_x} and Fig.~\ref{fig:w2_x}.
The information in Fig.~\ref{fig:allsimulation-clock} includes the
average, maximum, and minimum simulation wall clock times where each experiment was executed ten times. 
The results reveal a linear relation between the increase in the number of simulated jobs and the proportional increase in the simulation wall clock time consumed by the \mbox{two-level} simulator.



\section{Reproducibility of  This Work }
\label{sec:reprod}
To ensure reproducibility of this work, apart from the information in Section~\ref{sec:exp} about the workloads and the simulated platform considered in this work, the code of the proposed two-level simulator is developed under the LGPL license, and is available upon request from the authors. 
Under the same LGPL license, the developed Java-based tool used for extracting the jobs corresponding to the most intensive time period is also available upon request.
The Table~\ref{tab:rep1} summarizes the software and the hardware specifications of the platform on which the proposed two-level simulator has been compiled and executed.

\begin{table}
	\centering
	\captionsetup{font=footnotesize,labelsep=periodspace,singlelinecheck=false, justification=centering}
	\caption{Characteristics of the platform used to execute experiments }
	\label{tab:rep1}	
	\begin{tabular}{l|l}
		\multicolumn{2}{c}{\textbf{Software}}\\ \hline
		\begin{tabular}[c]{@{}l@{}}Operating System\end{tabular} & OS X 10.11.5 \\ 
		\begin{tabular}[c]{@{}l@{}}Required Libraries\\ For Build\end{tabular} & \begin{tabular}[c]{@{}l@{}}GridSim v.5Alea v.4\\ SimGrid v.3.14\end{tabular} \\ 
		Compilers & \begin{tabular}[c]{@{}l@{}}For SimGrid, clang v.7.3.0\\ For GridSim, javac v.1.8.0.91\end{tabular} \\ 
		\multicolumn{2}{c}{\textbf{Hardware}}\\ \hline
		Processor Model & Intel Core i7 \\ 
		Processor Frequency & 2.5 GHz \\ 
		RAM Size (DDR4) & 16 GB \\ 
	\end{tabular}
\end{table}


\section{Conclusion and Future Work}
\label{sec:conc}
With the growing complexity of modern and future HPC systems, parallelism becomes more massive and available at additional hardware levels. 
As a consequence, efficient exploitation and scheduling at these levels of parallelism is required.
It is, therefore, important and necessary to explore the relation between different levels of scheduling to enhance the performance and utilization of modern HPC systems as a whole and not only at individual scheduling levels. 
This work can be considered as an important first step in this direction. 
The proposed two-level simulation approach, connects simulators from two scheduling levels (BLS and ALS) and showed its validity to explore the relation between BLS and ALS.
Based on the proposed approach, a novel \mbox{two-level} simulator was proposed and successfully used to simulate combinations of three ALS and four BLS well-known algorithms from the literature.
The choice of ALS not only affects the performance of the applications for which it was employed, but also the performance of the chosen BLS for the other jobs in the batch. 
The proposed two-level simulator also generates execution traces in binary format.
This enables the visual analysis of the job execution at the batch level and task execution within each job at the application level. 
Such an analysis can help identify severe load imbalance and execution ``hotspots".
This type of insight is an important advantage for the development of future multi-level scheduling algorithms. 
The first evaluations of the proposed \mbox{two-level} simulator indicate its usefulness and scalable performance with the number of simulated jobs. 
It was able to simulate 10,000 jobs and their tasks at BLS and ALS levels in less than 110 seconds.

Further work is needed and planned to deepen the understanding of the relation between different levels of scheduling in modern large-scale HPC systems.
It is important to understand the root-cause behind certain BLS-ALS combinations being able to better absorb the effect of variable job and task lengths than others.
This understanding can benefit the design of \mbox{two-level} scheduling algorithms.
Further work is also needed to study additional combinations of more complex BLS and ALS. 

\section*{Acknowledgment}
This work is in part supported by the Swiss National Science Foundation in the context of the “Multi-level Scheduling in Large Scale High Performance Computers” (MLS) grant, number 169123. The authors acknowledge the contribution of Mr.~Aram Yesildeniz to the code that converts text-based SimGrid output to OTF2 traces.

\bibliographystyle{ieeetr}
\bibliography{mlsliterature.bib}

\begin{thebibliography}{10}

\bibitem{ahn2014flux}
D.~H. Ahn, J.~Garlick, M.~Grondona, D.~Lipari, B.~Springmeyer, and M.~Schulz,
  ``{Flux: A Next-Generation Resource Management Framework for Large HPC
  Centers},'' in {\em Proceedings of the 43rd International Conference on
  Parallel Processing Workshops (ICPPW)}, pp.~9--17, September 2014,
  Minneapolis, USA.

\bibitem{NpCompleteScheduling}
J.~Ullman, ``{NP-complete scheduling problems},'' {\em Journal of Computer and
  System Sciences}, vol.~10, no.~3, pp.~384--393, 1975.

\bibitem{azmi2011scheduling}
Z.~R.~M. Azmi, K.~A. Bakar, M.~S. Shamsir, W.~N.~W. Manan, and A.~H. Abdullah,
  ``{Scheduling Grid Jobs Using Priority Rule Algorithms and Gap Filling
  Techniques},'' {\em {International Journal of Advanced Science and
  Technology}}, vol.~37, pp.~61--76, 2011.

\bibitem{PBS}
R.~L. Henderson, ``{Job scheduling under the Portable Batch System},'' in {\em
  Proceedings of the 1st Workshop on Job Scheduling Strategies for Parallel
  Processing (JSSPP)}, pp.~279--294, April 1995, California, USA.

\bibitem{LSF}
M.~Q. Xu, ``{Effective metacomputing using LSF Multicluster},'' in {\em
  Proceedings of the 1st International Symposium on Cluster Computing and the
  Grid (CCGRID)}, pp.~100--105, May 2001, Brisbane, Australia.

\bibitem{blellochMultiscale}
G.~E. Blelloch, L.~Blum, M.~Harchol-Balter, and R.~Harper, ``{Multiscale
  Scheduling: Integrating Competitive and Cooperative Scheduling in Theory and
  in Practice}.'' http://lambda-the-ultimate.org/node/2337, 2007.
\newblock \mbox{[Accessed: 18-02-2017]}.

\bibitem{buyya2002gridsim}
R.~Buyya and M.~Murshed, ``{GridSim: A toolkit for the modeling and simulation
  of distributed resource management and scheduling for grid computing},'' {\em
  {Concurrency and Computation: Practice and Experience}}, vol.~14, no.~13-15,
  pp.~1175--1220, 2002.

\bibitem{casanova2014versatile}
H.~Casanova, A.~Giersch, A.~Legrand, M.~Quinson, and F.~Suter, ``{Versatile,
  scalable, and accurate simulation of distributed applications and
  platforms},'' {\em Journal of Parallel and Distributed Computing}, vol.~74,
  no.~10, pp.~2899--2917, 2014.

\bibitem{eschweiler2011open}
D.~Eschweiler, M.~Wagner, M.~Geimer, A.~Kn{\"u}pfer, W.~E. Nagel, and F.~Wolf,
  ``{Open Trace Format 2: The Next Generation of Scalable Trace Formats and
  Support Libraries},'' in {\em Proceedings of the International Conference on
  Parallel Computing (ParCo)}, pp.~481--490, April 2011, Ghent, Belgium.

\bibitem{simjava1998}
F.~Howell and R.~McNab, ``{A Discrete Event Simulation Library for Java},'' in
  {\em Proceedings of the 1st International Conference on Web-based Modeling
  and Simulation}, p.~6, January 1998, California, USA.

\bibitem{klusavcek2010alea}
D.~Klus{\'a}{\v{c}}ek and H.~Rudov{\'a}, ``{Alea 2: Job scheduling
  simulator},'' in {\em Proceedings of the 3rd International Conference on
  Simulation Tools and Techniques (ICST)}, p.~10, March 2010, Malaga, Spain.

\bibitem{dutot2016batsim}
P.-F. Dutot, M.~Mercier, M.~Poquet, and O.~Richard, ``{\mbox{Batsim:} A
  Realistic Language-Independent Resources and Jobs Management Systems
  Simulator},'' in {\em Proceedings of the 20th Workshop on Job Scheduling
  Strategies for Parallel Processing (JSSPP)}, p.~20, May 2016, Chicago, USA.

\bibitem{klusavcek2007alea}
D.~Klus{\'a}{\v{c}}ek, L.~Matyska, and H.~Rudov{\'a}, ``{Alea--Grid scheduling
  simulation environment},'' in {\em Proceedings of the 7th International
  Conference on Parallel Processing and Applied Mathematics (PPAM)}, p.~10,
  September 2007, Gdansk, Poland.

\bibitem{swfchapin1999benchmarks}
S.~J. Chapin, W.~Cirne, D.~G. Feitelson, J.~P. Jones, S.~T. Leutenegger,
  U.~Schwiegelshohn, W.~Smith, and D.~Talby, ``{Benchmarks and standards for
  the evaluation of parallel job schedulers},'' in {\em Proceedings of the 5th
  Workshop on Job Scheduling Strategies for Parallel Processing (JSSPP)},
  pp.~67--90, April 1999, San Juan, Puerto Rico.

\bibitem{srivastava2011enhancing}
S.~Srivastava, I.~Banicescu, F.~M. Ciorba, and W.~E. Nagel, ``{Enhancing the
  Functionality of a GridSim-based Scheduler for Effective Use with Large-Scale
  Scientific Applications},'' in {\em Proceedings of the 10th International
  Symposium on Parallel and Distributed Computing}, pp.~86--93, July 2011,
  Cluj-Napoca, Romania.

\bibitem{kruskal1985allocating}
C.~P. Kruskal and A.~Weiss, ``{Allocating independent subtasks on parallel
  processors},'' {\em IEEE Transactions on Software Engineering}, no.~10,
  pp.~1001--1016, 1985.

\bibitem{tang1986processor}
P.~Tang and P.-C. Yew, ``{Processor Self-Scheduling for Multiple-Nested
  Parallel Loops},'' in {\em Proceedings of the International Conference of
  Parallel Processing (ICPP)}, pp.~528--535, January 1986, Urbana, USA.

\bibitem{polychronopoulos1987guided}
C.~D. Polychronopoulos and D.~J. Kuck, ``{Guided self-scheduling: A practical
  scheduling scheme for parallel supercomputers},'' {\em IEEE Transactions on
  Computers}, vol.~100, no.~12, pp.~1425--1439, 1987.

\bibitem{hummel1992factoring}
S.~F. Hummel, E.~Schonberg, and L.~E. Flynn, ``{Factoring: A method for
  scheduling parallel loops},'' {\em Communications of the ACM}, vol.~35,
  no.~8, pp.~90--101, 1992.

\bibitem{caniou2008simbatch}
Y.~Caniou and J.-S. Gay, ``{Simbatch: An API for simulating and predicting the
  performance of parallel resources managed by batch systems},'' in {\em
  Proceedings of the European Conference on Parallel Processing (Euro-Par)},
  pp.~223--234, August 2008, Canary Island, Spain.

\bibitem{ciorba2012combined}
F.~M. Ciorba, T.~Hansen, S.~Srivastava, I.~Banicescu, A.~A. Maciejewski, and
  H.~J. Siegel, ``{A combined dual-stage framework for robust scheduling of
  scientific applications in heterogeneous environments with uncertain
  availability},'' in {\em Proceedings of the 21st International Heterogeneity
  in Computing Workshop (HCW) of the 26th Parallel and Distributed Processing
  Symposium Workshops \& PhD Forum (IPDPSW)}, pp.~193--207, May 2012, Shanghai,
  China.

\bibitem{hansen2014heuristics}
T.~Hansen, F.~M. Ciorba, A.~A. Maciejewski, H.~J. Siegel, S.~Srivastava, and
  I.~Banicescu, ``{Heuristics for Robust Allocation of Resources to Parallel
  Applications with Uncertain Execution Times in Heterogeneous Systems with
  Uncertain Availability},'' in {\em Proceedings of the 19th International
  Conference of Parallel and Distributed Computing (ICPDC) of the 25th World
  Congress on Engineering (WCE)}, p.~6, July 2014, London, UK.

\bibitem{AhmedSC16}
A.~Eleliemy, A.~Mohammed, and F.~M. Ciorba, ``{Extended abstract: Simulating
  Batch and Application Level Scheduling Using GridSim and SimGrid},'' Poster
  at the 28th International Conference for High Performance Computing,
  Networking, Storage, and Analysis (SC16), November 2016, Salt Lake City, USA.

\bibitem{feitelson2014experience}
D.~G. Feitelson, D.~Tsafrir, and D.~Krakov, ``{Experience with using the
  Parallel Workloads Archive},'' {\em Journal of Parallel and Distributed
  Computing}, vol.~74, no.~10, pp.~2967--2982, 2014.

\bibitem{parallelarchive2016}
D.~G. Feitelson, ``{Parallel Workloads Archive}.''
  http://www.cs.huji.ac.il/labs/parallel/workload/, 2005.
\newblock \mbox{[Accessed: 18-02-2017]}.

\bibitem{lublin2003workload}
U.~Lublin and D.~G. Feitelson, ``The workload on parallel supercomputers:
  Modeling the characteristics of rigid jobs,'' {\em {Journal of Parallel and
  Distributed Computing}}, vol.~63, no.~11, p.~18, 2003.

\bibitem{cirne2001model}
W.~Cirne and F.~Berman, ``{A model for moldable supercomputer jobs},'' in {\em
  {Proceedings of 5th International Parallel and Distributed Processing
  Symposium (IPDPS)}}, p.~8, April 2001, San Francisco, USA.

\bibitem{pfeiffer2008modeling}
W.~Pfeiffer and N.~J. Wright, ``{Modeling and predicting application
  performance on parallel computers using HPC challenge benchmarks},'' in {\em
  {Proceedings of the 13th International Parallel and Distributed Processing
  Symposium (IPDPS)}}, pp.~1--12, April 2008, Florida, USA.

\bibitem{knupfer2008vampir}
A.~Kn{\"u}pfer, H.~Brunst, J.~Doleschal, M.~Jurenz, M.~Lieber, H.~Mickler,
  M.~S. M{\"u}ller, and W.~E. Nagel, ``{The Vampir performance analysis
  tool-set},'' in {\em Proceedings of the 2nd International Workshop on
  Parallel Tools for High Performance Computing}, pp.~139--155, July 2008,
  Stuttgart, Germany.

\end{thebibliography}

\end{document}